\documentclass[apj]{emulateapj}
\usepackage{natbib}
\begin{document}

\title{High-Resolution Near Infrared Spectroscopy of HD~100546: II. Analysis of variable rovibrational CO emission lines}

\author{Sean D. Brittain}
\affil{Department of Physics \& Astronomy, 118 Kinard Laboratory, Clemson University, Clemson, SC 29634, USA; sbritt@clemson.edu}

\author{Joan R. Najita }
\affil{National Optical Astronomy Observatory, 950 N. Cherry Ave., Tucson, AZ 85719, USA; najita@noao.edu}

\author{John S. Carr}
\affil{Naval Research Laboratory, Code 7211, Washington, DC 20375, USA; carr@nrl.navy.mil}

\author{Joseph Liskowsky}
\affil{Department of Physics \& Astronomy, 118 Kinard Laboratory, Clemson University, Clemson, SC 29634, USA; jliskow@clemson.edu}

\author{Matthew R. Troutman}
\affil{Department of Physics \& Astronomy, University of Missouri - St. Louis, St. Louis, MO 63121, USA; troutmanm@umsl.edu}

\author{Greg W. Doppmann}
\affil{W.M. Keck Observatory, 65-1120 Mamalahoa Hwy, Kamuela, HI  96743, USA, gdoppmann@keck.hawaii.edu}

\begin{abstract}
We present observations of rovibrational CO in HD~100546 from four
epochs spanning January 2003 through December 2010. We show 
that the equivalent widths of the CO lines vary during this
time period with the v=1-0 CO lines brightening more than the
UV fluoresced lines
from the higher vibrational states.
While the
spectroastrometric signal of the hot band lines remains constant 
during this period,
the spectroastrometric signal of the v=1--0 lines varies substantially.
At all epochs, the spectroastrometric signals of the UV fluoresced  
lines are
consistent with the signal one would expect from gas in an  
axisymmetric disk.
In 2003, the spectroastrometric signal of the v=1-0 P26 line was  
symmetric and
consistent with emission from an axisymmetric disk.
However, in 2006, there was no spatial offset of the signal detected on the
red side of the profile, and in 2010, the spectroastrometric offset 
was yet more strongly reduced toward zero velocity.
A model is presented that can explain the evolution of the equivalent  
width of the v=1-0
P26 line and its spectroastrometric signal by adding to the system a  
compact source of
CO emission that orbits the star near the inner edge of the disk.
We hypothesize that such emission may arise from a circumplanetary  
disk orbiting
a gas giant planet near the inner edge of the circumstellar disk.
We discuss how this idea can be tested observationally and be
distinguished from an
alternative interpretation of random fluctuations in the disk emission.\end{abstract}

\keywords{stars: circumstellar matter, individual (HD 100546), protoplanetary disks}

\section{INTRODUCTION}
High-resolution near-infrared spectroscopy of rovibrational CO emission provides unparalleled
 views of gas in the inner disk around young stars \citep[e.g.][]{1996ApJ...462..919N, 2000prpl.conf..457N, 
 2003ApJ...589..931N, 2007prpl.conf..507N, 2001ApJ...551..454C, 2004ApJ...606L..73B, 
 2003ApJ...588..535B, 2007ApJ...659..685B, 2006ApJ...652..758G, 2008ApJ...684.1323P, 
 2011ApJ...733...84P, 2009ApJ...699..330S, 2011A&A...533A.112H}. Such observations serve 
 as a surrogate for direct imaging of this spatially unresolved region by spectrally resolving the 
 rotationally broadened emission lines.  If the stellar mass and disk inclination are known, the 
 spatial extent of the gas in Keplerian orbit can be inferred from the kinematic profile of the line. 
 Additional information about the spatial extent of the gas can be inferred by modeling the excitation.

In optically thick disks, the emission lines are formed in the warm optically thin atmosphere of 
the disk. If the gas in the atmosphere has a density of n$\rm_H$ $\gtrsim$ 10$^{12}$ cm$^{-3}$ 
and temperature of order 1000~K, the population of v$^\prime\ \geq$~1 vibrational levels 
excited by collisions can be significant and give rise to rovibrational emission lines. Such 
emission is commonly observed from the inner few AU of optically thick disks around classical 
T~Tauri stars \citep[e.g.][]{2003ApJ...589..931N, 2011ApJ...743..112S, 2009ApJ...699..330S} and 
Herbig Ae/Be stars 
\citep[e.g.][]{2004ApJ...606L..73B, 2007ApJ...659..685B}. In addition to collisional excitation, a 
source with a strong ultraviolet continuum  can fluoresce the gas \citep{2007ApJ...659..685B, 
2009ApJ...702...85B}. In the case of CO, the strongest electronic transitions fall near 1500~\AA.   
When CO is excited electronically, the molecule relaxes to the ground electronic state and 
populates excited vibrational levels. Molecules in these states give rise to hot band rovibrational 
emission lines ($\Delta$v=1 v$^{\prime} \geq 2$) such that the vibrational temperature of the gas 
reflects the diluted color temperature of the ultraviolet radiation field \citep{1980ApJ...240..940K}. 
The oscillator strength of the rovibrational transitions are a few $\times$10$^5$ weaker than the 
oscillator strengths of the electronic transitions. Thus the infrared rovibrational emission lines 
that arise from the electronic excitation of CO lines are optically thin. 

Curiously, the line luminosity 
of sources dominated by collisional excitation is comparable to that of sources dominated by UV 
fluoresced gas (e.g. Brittain et al. 2003). Thus the CO emitting area of sources dominated by 
UV fluorescence must be much larger than that of sources dominated by emission from 
collisionally excited gas. Indeed, rovibrational CO 
emission arising from UV fluorescence is observed to extend to $\sim$100~AU (e.g. HD~141569, 
Goto et al. 2006, Brittain et al. 2007; HD~100546, van der Plas et al. 2009, Brittain et al. 2009; 
HD~97048, van der Plas et al. 2009; Oph~IRS~48, Brown et al. 2012), whereas collisionally 
excited CO arises from a much smaller area (from radii within $\sim 1$\,AU in T Tauri disks). 

In addition to modeling the spectral line profile and excitation of the gas, information about the spatial 
distribution of the gas can be determined using spectroastrometry \citep{2008LNP...742..123W}. 
Spectroastrometry is a relatively new technique that takes advantage of high-resolution spectrometers 
with excellent imaging quality (Bailey et al. 1998; Takami et al. 2001; Takami et al. 2003; Acke \& 
van~den~Ancker 2006; Pontoppidan et al. 2008, 2011; Brittain et al. 2009; van der Plas et al. 2009). 
The projected velocity of gas in a rotating disk varies as a function of distance from the star and orbital 
phase. Regions of common projected velocity map out extended loop-like structures. Thus the centroid 
of the point spread function of each velocity channel of a resolved emission line formed in the disk is 
offset from the position of the star. By simultaneously fitting the kinematic line profile and spectroastrometric 
signature of the line, one can provide an independent constraint on the stellar mass and disk inclination 
(Pontoppidan et al. 2008) and identify non-Keplerian emission signatures \citep[e.g.][]{2011ApJ...733...84P}. 

\object{HD~100546} - a nearby  $\sim$10 Myr, 2.4~M$_{\sun}$, isolated Herbig Be star (HBe; B9Ve) - 
is a particularly exciting source to which to apply this new tool. HD~100546 has been well studied 
because of its proximity to our Solar System (d=103$^{+7}_{-6}$ pc; van den Ancker 1997). The mass of 
dust in the disk is $5\times10^{-4} \rm M_\sun$ assuming that the dust is 50K and optically thin (Henning 
et al. 1998). While the absolute mass of the gas in the disk is not known, it is a rich source of molecular 
emission lines (Pani{\'c} et al. 2010; Sturm et al. 2010; Thi et al. 2011; Bruderer et al. 2012; Liskowsky et al. 2012). 
Indeed, Bruderer et al. (2012) model the rotational emission line spectrum of CO and find that the high-J 
lines (e.g. J=30-29) originate from $\sim$20-50~AU, while the mid-J lines (e.g. J=16-15) arise mainly from  
$\sim$40-90~AU. HD~100546 is also a transition object - a star whose disk is optically thin in the inner 
region and optically thick in the outer region. The optically thin inner region of the disk extends to $\sim$ 
10~AU (Bouwman et al. 2003; Grady et al. 2005) and is devoid of CO and OH inward of this radius 
(van der Plas et al. 2009; Brittain et al. 2009; Liskowsky et al. 2012); however, there is evidence of residual 
atomic gas (\ion{O}{1}) in the  inner region (Acke \& van den Ancker 2006) and ongoing stellar accretion 
(Guimar{\~a}es et al. 2006). 

As described in Brittain et al. (2009), measurement of the half width at zero intensity (HWZI) of the 
emission lines indicates that the rovibrational 
CO emission extends into 13$\pm$6 AU. The vibrational temperature ($\sim$6000~K) is higher than the 
rotational temperature of the gas ($\sim$1000~K), the tell tale signature of UV fluoresced gas. The $^{13}$CO 
lines and $^{12}$CO lines from v$^{\prime} \geq$2 were shown to be populated by UV fluorescence while 
the $^{12}$CO v=1--0 lines were shown to be excited by a combination of UV fluorescence and collisional 
excitation. Both $^{12}$CO and $^{13}$CO lines are excited over a column that corresponds to $\tau\sim1$ 
in the UV transitions that induce the fluorescence.  This translates into a physical column that is much larger 
for $^{13}$CO than $^{12}$CO due to its much lower abundance. The optically thin gas is warm and can 
populate v$^{\prime}$=1 collisionally as well. However, the ratio of this contribution scales with the relative 
abundance of the two isotopologues such that the collisional contribution to the $^{12}$CO lines is nearly 
two orders of magnitude larger than the collisional contribution to the $^{13}$CO lines. The line profile and 
flux was reproduced by assuming the emitting gas extends from 13-100 AU \citep{2009ApJ...702...85B}.  

In this paper we present multi epoch high resolution rovibrational CO emission line spectra of  \object{HD~100546}  
and compare the spectra taken in four epochs. In a companion paper (Liskowsky et al. 2012; hereinafter Paper I), 
we considered the rovibrational OH emission and rovibrational CO hot band emission from HD100546. Here we 
complement Paper I by focusing on the interpretation of 
the 
CO v=1--0 emission. We show that the equivalent width, 
line profile, and spectro-astrometric signal of the CO emission lines vary over the nearly eight year span of our 
observations. We model the spectro-astrometric signature of the emission lines and discuss the possibilities for 
the origin of this variability.

%%%%%%%%%%%%
% OBSERVATIONS %
%%%%%%%%%%%%
\section{Observations}
We acquired high-resolution (R=50,000), near-infrared spectra of HD~100546 on  
January 14, 2006, December 21-22, 2008, and December 23, 2010 using PHOENIX at the Gemini 
South telescope \citep{2003SPIE.4834..353H,2000SPIE.4008..720H,1998SPIE.3354..810H}. 
The 4 pixel slit (0\arcsec.34) was used for all observations. HD~100546 was acquired by first taking 
a $K$-band image and centering the star near the nominal position of the center of the slit. 
Next an $M$-band image was acquired with the slit in the beam. Then the slit was taken out 
and a series of short exposures were taken. The telescope was moved a few arcseconds 
and another series of images were taken. These images were differenced and the position 
of the star was determined. The telescope was then moved so that the star was centered on the 
slit. In addition to the data we have taken, the spectrum acquired on January 7, 2003 was 
taken from the archive. The data were also acquired with the 0$\arcsec$.34 slit providing 
a resolution of R=50,000. The acquisition procedure is not recorded, but it is likely the standard 
procedure described above was followed.

Observations centered near 2034~cm$^{-1}$ were acquired in 2003, 2006, and 2010 to cover the 
v=1--0 P26 line (hereafter we will refer to this transition as P26) and numerous hot band lines. Observations 
centered near 2144 cm$^{-1}$ were acquired in 2006 and 2008. The spectra overlap from 2029~cm$^{-1}$ 
to 2035~cm$^{-1}$ (Fig. 1) and from 2141~cm$^{-1}$ and 2147~cm$^{-1}$ (Fig. 2).   The observations 
made in 2006 have been presented in Brittain et al. (2009). The other observations have not been 
reported previously.

 The seeing in the $M$-band during these epochs ranged from 0$\arcsec$.4-0$\arcsec$.8 and the airmass of the
 observations ranged from 1.3-1.6. A 
 summary of observations is presented in Table \ref{tab:observations}. Most of the data were 
 taken with the slit in its default position angle of 90$\degr$ east of north (all position angles 
 quoted in this paper are measured east of north unless otherwise noted) with the exception 
 of the observations acquired in 2008. These observations were taken with the slit rotated to 
 37$\degr$ and  127$\degr$ - near the apparent semi-minor and semi-major axes respectively, 
 i.e., along the semi-minor and semi-major axes as projected on the sky. An antiparallel 
 observation near the apparent semi-major axis (PA=307\degr) was acquired as well.

Observations in the $M$-band are dominated by a strong thermal background. Therefore, an ABBA nod pattern
 between two positions separated by $\sim 5\arcsec$ is used to cancel the thermal continuum to first order. 
 The extent of the CO emission is $\sim$0.7$\arcsec$ in each direction (Brittain et al. 2009), thus there is 
 no contamination between the beams. The
  scans are flat fielded, cleaned of hot and dead pixels, including cosmic ray hits, and then combined in the
   sequence (A$_1-$B$_1-$B$_2+$A$_2$)/2. Because the spectra are curved along the detector, they are first
 rectified by finding the centroid of the point spread function (PSF) in each column. A third degree polynomial was 
 fit to the centroids and used to determine the shift of each column to a common row. 

In principle, the spectra can also be tilted along the slit. To check this, we compared the sky emission spectrum at 
the bottom of the slit to the spectrum at the top of the slit. The offset was less than half of a pixel along the length of 
the slit. A sky emission spectrum generated by the Spectral Synthesis Program \citep{1974JQSRT..14..803K}, 
which  accesses the 2000HITRAN molecular database \citep{2003JQSRT..82....5R} was used to measure the
shift of each row.  A first degree polynomial was fit to the central wavenumber of each row to measure the shift. 
Each row was then rectified in the spectral direction. 

\begin{figure}
    \begin{center}
	\includegraphics[scale=.5]{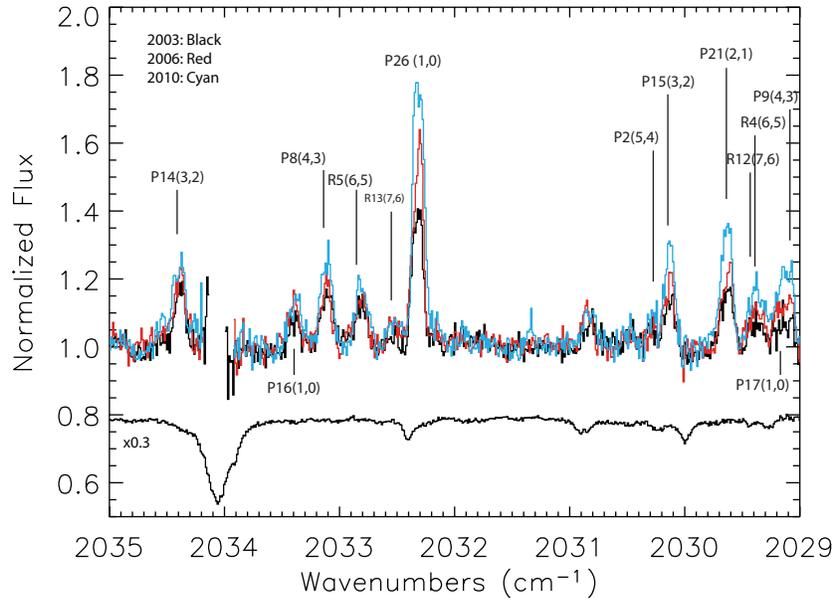}
	\caption[CO Spectra]{Observations of rovibrational CO emission from HD~100546 spanning 2003 to 
	2010. The spectra of HD~100546 were normalized with a telluric standard and regions with a 
	transmittance less than 50\% are omitted. The  $^{12}$CO emission lines are labeled above the 
	spectrum and the $^{13}$CO  emission lines are labeled from below. The lines used for constructing 
	the average profile are given in Table 2. The telluric standard from 2006 is plotted below the 
	spectrum. The change in the equivalent width of the hot band lines is likely due to the change in the 
	continuum emission. }
	\label{fig:fig1}
   \end{center}
\end{figure}

\begin{figure}
    \begin{center}
	\includegraphics[scale=.5]{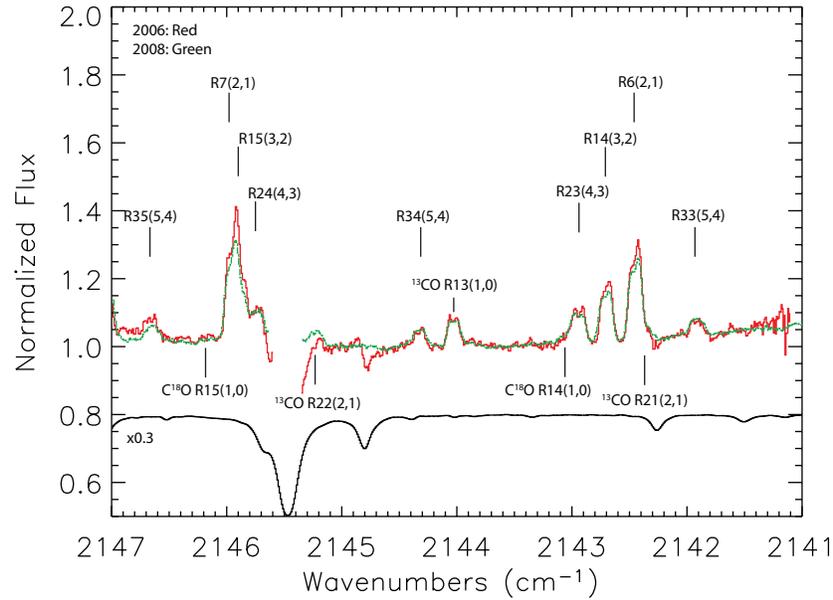}
	\caption[CO Spectra]{Observations of rovibrational CO emission from HD~100546 in 2006 (red) and 
	2008 (green). The spectra of HD~100546 were normalized with a telluric standard and regions with a 
	transmittance less than 50\% are omitted. The data are offset for clarity.  The hot band lines used to 
	construct the average profile are presented in table 2. The telluric standard from 2006 is plotted 
	below the spectrum.}
	\label{fig:fig2}
   \end{center}
\end{figure}

To capture all of the CO emission, a 2$\arcsec$ window was extracted to generate the one-dimensional 
spectrum. The wavelength calibration was achieved by fitting an atmospheric transmittance model generated by 
the Spectral Synthesis Program. Each spectrum is then ratioed to a standard star observed at a similar airmass to 
remove telluric absorption lines. Areas where the transmittance is below 50\% are omitted (Figs. \ref{fig:fig1} \& 
\ref{fig:fig2}).

%%%%%%%%%
% RESULTS
%%%%%%%%%
\section{Results}%Section 3
\subsection{Spectral line comparison} %Section 3.1
The spectra observed in 2003, 2006, and 2010 overlap from 2029 cm$^{-1}$ to 2035 cm$^{-1}$ (Fig. \ref{fig:fig1}) 
and the spectra observed in 2006 and 2008 overlap from 2141 cm$^{-1}$ to 2147 cm$^{-1}$ (Fig. \ref{fig:fig2}). In 
each spectrum, $^{12}$CO emission lines arising from the upper vibrational levels v$^{\prime}$=1,2,3, 4, 5, and 6 
are detected. $^{13}$CO emission lines arising from the upper vibrational levels v$^{\prime}$=1 \& 2 are also 
detected. The radial velocity of the source we infer from the Doppler shift of the hot band emission lines is 
v$_\sun$=+16$\pm$1~km~s$^{-1}$ which we take to be the system velocity. This is 
consistent with the radial velocity of the gas giving rise to the rotational CO lines observed at millimeter 
wavelengths (Pani{\'c} et al. 2010) and optical atomic lines (Acke et al. 2005). 

\begin{figure}
    \begin{center}
	\includegraphics[scale=.5]{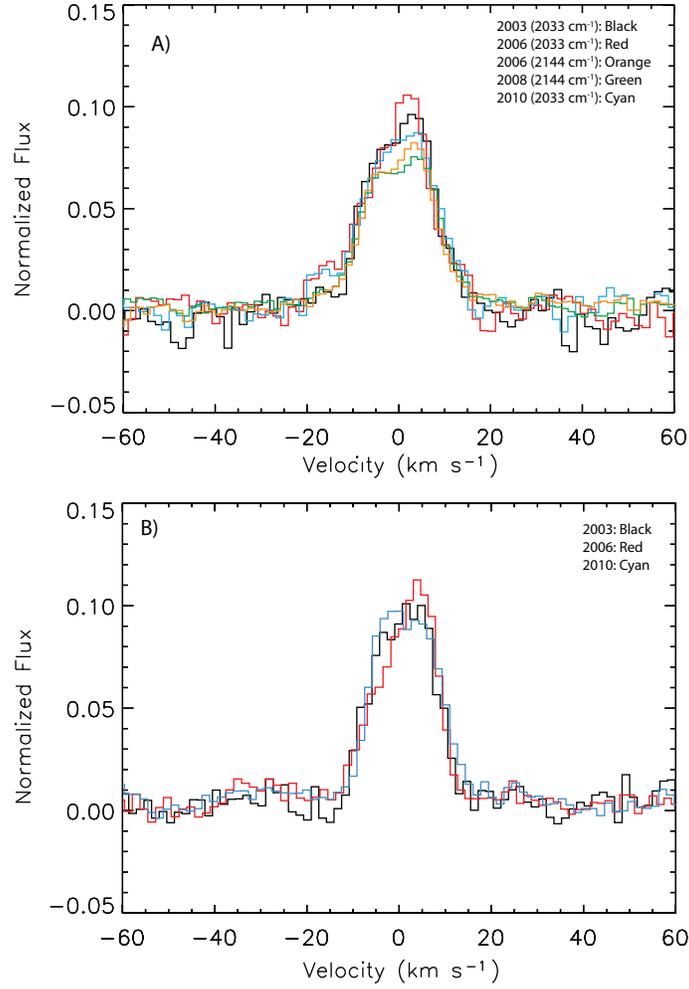}
	\caption[Averaged Spectra]{The average profile of hot band rovibrational CO emission lines near 
	2033 cm$^{-1}$ and 2144 cm$^{-1}$  (A) and the line profile of the $^{12}$CO v=1-0 P26 transition 
	(B). All of the profiles have been normalized to a common equivalent width to highlight the variation of 
	the continuum. The average hot band emission line is asymmetric in all three epochs though the 
	average profile of the hot band lines from the order centered near 2033 cm$^{-1}$ observed in 2006 
	(solid red spectrum in panel A) shows a larger red excess. The average line profile of the hot band 
	lines can be fit with a 25\% contribution from an eccentric inner annulus and a 75\% contribution from 
	a circular outer disk (Paper I). The P26 line was normalized  shows a significant line shape variation. 
	In 2003 (black), the line showed a slight red excess. In 2006 (red), the line showed a pronounced 
	excess on the red side, and in 2010 (cyan), the line showed a slight excess on the blue side of the 
	line.}
	\label{fig:fig3}
   \end{center}
\end{figure}

{\it Line Asymmetry.}
To enhance the signal to noise of the weaker line profiles, we have averaged the unblended lines whose 
excitation is dominated by UV fluorescence in each order on each date (Table 2). Figure 3a shows the resulting 
average normalized profiles of lines in the orders centered on 2033 cm$^{-1}$ and 2144 cm$^{-1}$. We find that 
there is a persistent asymmetry in the lines.  The profile of the stronger P26 line is also asymmetric, but 
in contrast to the hot band lines, the shape of the line varies considerably (Fig. \ref{fig:fig3}b).  

To quantify the asymmetry we compare the ratio of the red side 
of the line (from v=0 km s$^{-1}$ to +5~km~s$^{-1}$) to the blue side of the line (from v=0 km s$^{-1}$ to $-
$5~km~s$^{-1}$).  This ``asymmetry ratio'' for the UV fluoresced lines is $\sim$1.1, with the 
exception of the average line profile of the lines observed near 2033~cm$^{-1}$ in 2006. The asymmetry ratio 
for these lines is 1.18 on this date (Fig. 3a). The asymmetry of the P26 line flips from 2003 to 2010. The 
asymmetry ratio of the line is 1.06, 1.37, and 0.97 in 2003, 2006, and 2010 respectively 
(Fig. 3b).  

The red excess seen in the average of the hot-band emission 
lines is the opposite of what is observed for the rovibrational OH 
emission from this source. As shown in Paper I, the OH line shows a prominent blue excess that can be explained if 
the OH arises from an annular inner rim with an eccentricity of 0.18. In Paper I, we also show that we can fit the 
average profile of the UV fluoresced CO lines reasonably well if most of the CO comes from the outer circular 
disk rather than the eccentric inner rim. 

The asymmetry of the hot band lines also varies with vibrational level. 
Close inspection 
of the spectrum centered near 2144~cm$^{-1}$ reveals that the v=2--1 R6 line is more asymmetric than the v=3--2 R14 
line (though there is a weak contribution on the red wing of the v=2--1 R6 line from the $^{13}$CO v=2--1 R21 line). 
The asymmetry of the v=3--2 R14 line is larger than that of the v=4--3 R23 line, and the v=4--3, v=5--4 and $^{13}$CO 
lines are symmetric. The origin of this trend is not entirely clear. One possibility is that the relative contribution of the 
eccentric inner rim to the emission lines is not uniform across the observed vibrational levels. In our modeling we found that we
needed to add a collisional component to our model to fit the v=1--0 lines. It is possible that this collisional component also 
contributes some to the v=2--1 and v=3--2 lines. For CO in LTE at a temperature of 1400~K, the population 
of v$^{\prime}$=2 and v$^{\prime}$=3 relative to v$^{\prime}$=1 is 10\% and 1\% respectively. To test this 
possibility, high signal to noise spectra of CO across a broad range of energy levels in a single epoch is 
necessary and is the subject of ongoing analysis.

{\it Equivalent Width Variation.} 
As shown in Figures 1 and 2, 
the equivalent width of the CO lines varied over the span of the observations.  
The average equivalent width of the hot band lines increased by a factor of 1.14$\pm$0.12 from 2003 to 
2006, whereas the P26 line brightened by a factor of 1.44$\pm$0.06. From 2006 to 2008, the hot band 
lines dimmed by a factor of .93$\pm$0.03. From 2003 to 2010, the hot band lines brightened by a factor 
of 1.53$\pm$0.13 and the P26 line brightened by a factor of 2.18$\pm$0.07 (Table 3).

Generally, the equivalent width variation of emission lines can be due to a varying line luminosity or a varying 
continuum. 
The line luminosity of the rovibrational CO hot band lines resulting from UV fluorescence depends on the 
FUV luminosity of the source, the flaring of the disk, and velocity dispersion of the gas. 
In the far ultraviolet (FUV) spectrum of HD100546, 
acquired with the $International\ Ultraviolet\ Explorer$ in 1993 and 1995 (Grady et al.\ 1997), 
the continuum varied by less than 5\% over this timespan. 
Thus, the FUV continuum does not appear to be variable on time 
scales of two years and it is unlikely that disk flaring and velocity dispersion of the gas would vary globally by $\sim
$50\% over the span of our observations. 
We therefore conclude that it is unlikely that the flux of the UV fluoresced CO 
lines varies as dramatically as indicated by the variation in the equivalent width. 

It seems more likely that the fluctuation of the equivalent width of the hot band CO lines is due to 
a varying $M$-band 
continuum. This would imply that the $M$-band flux has varied by $\sim$0.4 mag over a span of eight years. 
Morales-Calder{\'o}n et al. (2011) find that about 5\% of class II young stars in Orion showed  variation in the $M-
$band $\geq 0.5$\,mag over a 40 day span. Similarly, Muzerolle et al. (2009) find that the transitional disk LRLL 31 
varied by as much as 30\% over time periods as short as a week. Espaillat et al. (2011) find similar results for a 
sample of 14 transitional and pre-transitional T~Tauri stars. 

We can assess the degree of the NIR variability of
HD~100546 by comparing the relative magnitude of the
acquisition images of HD~100546 and the telluric standards.
We do not have $M$-band acquisition images of
our targets on all dates, but we can gain some insight
into the degree of variability by comparing the images we
do have.

In 2003 and 2010, we imaged HD~100546 and standards in a
narrow-band filter in the $K$-band. 
These images 
show that HD 100546 was 0.5$\pm$0.15 mag fainter
in the $K$-band in 2010 compared to 2003.
In 2008, $M$-band acquisition
images of HD~100546 and HR~4798 allowing us to
estimate that $M$=3.82 mag. This is consistent with the
reported $M$-band magnitude reported in the literature
($M$=3.80; Malfait et al. 1998).
In 2010, we estimate the
$L$-band magnitude was 4.92. This is 50\% fainter than
the reported brightness between 1988 and 1992 ($L$=4.08;
Malfait et al. 1998).

In summary, HD 100546 was
0.5mag fainter in the $K$-band in 2010 than in 2003,
similar to the change in the $M$-band required to explain
the change in the equivalent width of the hot band
CO lines between these two epochs.
In addition, HD~100546 was 0.84mag fainter in the $L$-band
in 2010 than in 1990.
While not conclusive, these measurements do suggest that
known variations in the infrared continuum in HD~100546 are
sufficient to explain the variation in the equivalent width
of the hot band CO lines. 

The origin of the variability is uncertain. Modeling of the SED 
indicates that the majority of the NIR flux arises from small grains 
in the inner 1~AU of the disk.
These are possibly formed by the grinding of planetesimals 
in this inner region. (Bouwman et al. 2003). It is plausible that the 
variability is due to the variation of the abundance of small grains 
in this region.

\subsection{Spectro-astrometric analysis}%Section 3.2
In addition to comparing the spectral profile and equivalent width of the lines, we can also compare the spatial 
distribution of the emission by making spectro-astrometric measurements of our data. The large
angular extent of the rovibrational CO emission from HD~100546 
($\sim$1.5$\arcsec$; Brittain et al. 2009) is an asset in this regard. 
In a spectroastrometric measurement, when the slit of the 
spectrograph is aligned with the apparent semi-major axis of an inclined, intrinsically circular disk, the 
spatial centroid of the redshifted and blueshifted 
sides of the line will be spatially offset. When the slit is aligned with the apparent semi-minor axis of the disk, 
the center of the 
redshifted and blueshifted sides of the disk will be centered on the PSF. We present continuum subtracted 
images of our spectra centered near 2144~cm$^{-1}$ for four slit PAs. These position velocity diagrams show the 
extension of the emission lines shift with the rotation of the slit (Fig 4).  

To measure the spectro-astrometric signal of the CO emission lines, we follow the method outlined by \citet
{2008ApJ...684.1323P}.  The spatial center of light of the continuum is determined by averaging the columns of the 
spectrum that show no emission features or absorption lines whose transmittance is less than 50\%. The centroid of 
this average continuum PSF is used as the fiducial center of light of the system along the slit, $x_0$. 

For each 
column of the spectrum, the centroid of the PSF is measured,
 \begin{equation}
	X_{\tilde{\nu}}=C\frac{\sum_i(x_i({\tilde{\nu}})-x_0)F_i({\tilde{\nu}})}{\sum_i F_i({\tilde{\nu}})}.
\end{equation}
For a given column designated by $\tilde{\nu}$, the vertical position of a pixel in a given row $i$ is designated $x_i$ 
and $F_i({\tilde{\nu}})$ is its flux. 
$C$ is the correction factor for the fraction of the PSF excluded from the window over which the signal is calculated. 
For each column of the spectrum, we measure the centroid of the flux within the FWHM, so we multiply by a 
correction factor of 1.25 to account for the missing flux  \citep{2008ApJ...684.1323P}. The shift of the centroid is 
measured in pixels, and each pixel subtends 0$\arcsec$.085. 

The center of light of a particular velocity channel of the spectrum depends on both the spatial extent of the line 
emission and the fraction of light that is extended (in other words the line to continuum ratio). Since we are 
measuring the spatial displacement of the line + continuum, a large continuum flux will dilute the spatial 
displacement of the line emission. We account for this when we calculate the spectro-astrometric signals 
from our spectral synthesis code by including the continuum (see section 3.3 for further details). A summary of 
of our investigation of potential sources of artifacts is presented in the appendix.
 
The data collected in 2003, 2006, and 2010 were not optimized for spectroastrometry, thus only the brightest line 
shows a significant signal in our data (Fig. 5).  The star was observed in the default position angle of the instrument 
(PA=90\degr east of north) while the apparent position angle of the disk is 138\degr$\pm$4\degr (Quanz et al. 
2011).  The data acquired in 
December 2008 were observed with the slit turned to 37$\degr$, 127$\degr$, and 307 $
\degr$. The standard deviation of our measurement of the centroid of the continuum ranges from  1.9-4.5 
milliarcseconds  in these epochs (Table 1). 

In 2003, the center of light of the PSF of the columns containing the P26 line was offset relative to the centroid of 
columns only containing continuum by as much as $\sim$15~$mas$ (Fig. 6a). The average shift of the columns 
containing hot band lines was also detected and was offset by as much as 7~$mas$ (Fig. 6d). The relatively symmetric 
offset (in one spatial direction for the red side of the profile and the opposite spatial direction for the blue side of the 
profile) is expected for gas arising from an axisymmetric rotating disk (Pontoppidan et al. 2008). 

In 2006, we detect 
the offset of the blue side of the P26 line (Fig. 6b), but we do not have the sensitivity to measure the offset of the hot 
band lines near 2034 cm$^{-1}$ (Fig. 6e). We do however measure the spectro-astrometric offset of the hot band 
lines observed near 2144~cm$^{-1}$ on the same night (Fig. 6h). The spectro-astrometric signal of the lines is 
consistent with gas in a axisymmetric Keplerian orbit, consistent with the interpretation given for these features in 
Paper I.

In 2008, we re-observed HD~100546 centered at 2144~cm$^{-1}$, however, the slit was aligned near the apparent 
semi-major axis and apparent semi-minor axis rather than in the default position of 90$\degr$ (Fig. 6g, i; Fig. 7). The average center 
of light of the hot band emission lines showed a very small offset (a maximum of 2.1$\pm$0.8 mas across the line) 
when the slit was nearly aligned with the apparent semi-minor axis (Fig. 6g) and a much larger offset (a maximum 
of 9.6$\pm$0.7 mas across the line) when the slit was aligned near the apparent semi-major axis (Fig. 6i). 

Finally, a brief follow up observation of the P26 line was 
acquired in December 2010 with the slit in the default 90$\degr$ position angle (Fig. 6c). As in 2006, the hot band 
lines revealed the tell tale offset of gas in a rotating axisymmetric disk while the P26 line only showed an offset on the blue side of 
the line. Curiously, the offset of the blue side of the line was only modestly larger than the offset observed in 2003 even though the 
line to continuum contrast had more than doubled. The minimal spatial extent of the red side of the P26 line in 2006 and 
2010 is indicative of non-axisymmetric structure in the disk. 

\subsection{Calculation of spectro-astrometric signals from spectral synthesis}%Section 3.3
In order to facilitate the interpretation of the spectro-astrometric measurements of the CO emission lines, we 
calculate the spectro-astrometric signal  of the lines generated by our modeling of the CO emission (Brittain et al. 
2009). We model the excitation of  CO in a flared, axisymmetric disk with a cleared inner region (within 13 AU) that 
is illuminated by the ultraviolet flux from the central star (M$_\star$=2.4$\pm$0.1M$_\sun$; van den Ancker et al. 
1997). The disk is assumed to be in Keplerian rotation with an inclination of 47$\degr$ and a position angle of 138$
\degr$, parameters that we adopt based on the recent Polarimetric Differential Imagery of HD~100546 (Quanz et al. 2011; see 
also Pantin et al. 2000; Augereau et al. 2001; Grady et al. 2001; Liu et al. 2003; Leinert et al. 2004; Ardila et al. 
2007). 

As in our 2009 paper, the temperature of the gas is described by the power-law $T(r)=1400~K(R/13~AU)^
{-0.3}$, and the CO is excited by a combination of UV fluorescence and 
collisions with hydrogen. The UV flux enters 
the flared disk at a relatively shallow angle (5$\degr$ in the inner region and increasing to 10$\degr$ at 100~AU) 
that is calculated assuming the disk is in hydrostatic equilibrium, vertically isothermal, and the scale height of the 
gas is set by the temperature described by the power law above. The inner edge of the disk has a scale height of 
3.5AU (Bouwman et al. 2003) and is normal to flux arriving from the star, thus a disproportionate amount of the flux 
arises from this inner rim ($\sim$15\%). The amount of CO emission arising beyond 100~AU is 
insignificant, consistent with the spatial extent of the emission we observe. 

Previously, we found that the v=1--0 CO lines calculated by our model were too weak if we entirely ignored 
collisional excitation. By adding in collisions with hydrogen we found that we could reproduce the line profiles and 
line fluxes with this model by adopting a density distribution of $\rm n(H)=10^{11} cm^{-3}(R/13~AU)^{-1}$. While 
the v=1--0 lines were weaker in the 2003 data than in the 2006 data set, we found that we still could not entirely 
ignore collisional excitation and found that we could reproduce the v=1--0 lines by adopting a hydrogen density 
described by a power-law with a slightly lower fiducial density, $\rm n(H)=2.5\times10^{10} cm^{-3}(R/13~AU)^{-1}$.

To obtain the spectro-astrometric signal from this model, we record the line luminosity,  projected velocity, orbital
phase, and projected distance from the star of every resolution element in the disk. The shift of the centroid of the 
PSF is then computed for the position angle of 
the observation. As shown by the model results, the spectroastrometric measurement of the P26 line observed in 
2003 and the hot band lines in all epochs 
is consistent with gas in an axisymmetric disk (Fig 6a, d-i). The spectro-astrometric 
signals of the 
P26 line observed in 2006 and 2010 are not (Fig. 6b and c). The red side of the spectro-astrometric signals from the 
P26 line observed in 2006 and 2010 is not extended. 

In principle, this could be caused by two effects. Either the redshifted disk 
emission at large radii has gone away or 
extra redshifted emission has arisen such that the center of light has shifted towards the center of the PSF. If the 
former had occurred, the low velocity emission
on the red side of the emission line should have decreased in strength which is not observed. 
Rather, the equivalent width of the emission
line has increased significantly relative to the increase of the fluoresced lines. If excess emission is instead the 
cause, the 
excess emission must be non-axisymmetric in order to decrease the extent of the red side of the spectro-astrometric 
signal. If we use a simple prescription to place the spectra on a flux scale, the excess emission can be attributed to 
a point source located near the star, as we show in the next section.

\subsection{Modeling the variations of the spectro-astrometric signals}%Section 3.4

In section 3.1 we concluded that the most likely explanation for the 
changes in the equivalent widths 
of the UV fluoresced lines is the  variation of the continuum. We therefore assume 
that the fluxes of the UV fluoresced lines are constant 
and scale the continuum subtracted spectra so that the UV fluoresced 
lines match. With this scaling, there is an increase in the flux of the P26
line in 2006 and 2010 relative to the 2003 spectrum.
We also showed in section 3.3 that the spectro-astrometric signal
from both the P26 line and the UV fluoresced lines in 2003 
can be reproduced by emission from an axisymmetric disk (Fig. 6a).
We therefore adopt the 2003 data as the fiducial spectrum of the
disk emission.

To examine whether the excess P26 line emission can
account for the variation of spectro-astrometric signals
observed in 2006 and 2010,
we subtract the scaled 2003 spectrum from
the scaled 2006 and 2010 spectra.
The difference spectra shown
in Figure 8 show the excess emission that results.
The excess v=1--0 emission observed in 2006 is redshifted
+6 $\pm$ 1 km s$^{-1}$ with respect to the system velocity and
is singly peaked. The excess v=10 emission observed in
2010 is shifted 1$\pm$ 1 km s$^{-1}$ with respect to the system
velocity and is double peaked.

If we assume that the continuum in 2003 was $1.5 \times
10^{-12}$ erg/s/cm$^2$/cm$^{-1}$ (based on the M-band magnitude
(Malfait et al. 1998), the flux of the P26 line was 6.8$\times$10$^{-14}$ erg~s$^{-1}$~cm$^{-2}$ and 
the excess emission
of the P26 line during the 2006 and 2010 observations 
was $\rm 1.5\times10^{-14}~erg~s^{-1}~cm^{-2}$ (a 22\% increase) and
$\rm 2.9\times10^{-14}~erg~s^{-1}~cm^{-2}$  (a 43\% increase) respectively. These fluxes are 
comparable to the fluxes of the P26 lines observed in classical T Tauri stars 
in Taurus (Najita et al. 2003).
The increase of the scaled equivalent width of the P26
line, the Doppler shift of the excess emission, and the  
FWHM of the excess emission are summarized in Table 4.

If the CO that produces the excess P26 line emission has a temperature 
of 1400~K, an intrinsic line width of 3~km~s$^{-1}$, and $\tau_0$=1, its emitting area is $
\sim$0.1AU$^2$. The absolute flux calibration is uncertain, but this analysis provides a rough idea of the scale
of the emitting area.

We model this extra source of emission as a point source, placing it near the inner edge of the disk ($\sim$13~AU) 
and giving the emission lines a shape similar to the residual (approximated as a Gaussian in 2006 and double 
Gaussian in 2010; Fig. 8). The FWHM of the synthetic Gaussian line used to approximate the excess flux in 2006 is 
6~km~s$^{-1}$, its Doppler shift is +6~km~s$^{-1}$, and its flux is $\rm 1.5\times10^{-14}~erg~s^{-1}~cm^{-2}$. The 
synthetic double Gaussian line used to approximate the excess flux in 2010 has a FWHM of 12~km~s$^{-1}$, a 
Doppler shift of $-1$~km~s$^{-1}$, and flux of $\rm 2.9\times10^{-14}~erg~s^{-1}~cm^{-2}$.  

If the excess arises from a source in a circular Keplerian orbit around 
the star that is coplanar with the disk, then to obtain the observed 
velocity centroid, the source of the excess emission would be 
located 6$\degr^{+8\degr}_{-10\degr}$ east of north in 
2006. Five years later, December 2010, the position of the excess 
emission is  54$\degr^{+7\degr}_{-6\degr}$ east 
of north (Fig. 9). We also find that the position angle of the 
source of the emission in 2003 would be $-30\degr^{+8\degr}_{-10\degr}$ 
east of north- just hidden from view by the large inner wall of the disk.  
Adding the excess flux to our model at these phases and velocities and 
recalculating the spectro-astrometric signal, 
we find that we reproduce the observed signal (Fig.~10).

%%%%%%%%
% ANALYSIS
%%%%%%%%
\section{Discussion}%section 4
What physical entity could be responsible for this excess CO v=1--0 
emission that varies in velocity centroid, velocity width, and emission strength? 
One possibility is that the excess CO emission we see, 
arises from random fluctuations in the disk. 
Fluctuations of this kind may arise from 
the magnetorotational instability (Balbus \& Hawley 1991, 1998; 
Hawley \& Balbus 1991).  In driving turbulence 
(and angular momentum transport) in disks, the instability 
is believed to produce 
fluctuations in density 
(e.g., as shown in detailed MHD simulations such as Flock et al. 2011), 
as well as disk height and localized 
heating around current sheets (Hirose \& Turner 2011). 
These effects may cause significant spatial inhomogeneities 
in the CO emission from the disk which would also vary with time. 
Detailed theoretical models are needed to explore this possibility. 

If fluctuations of this kind are responsible for the time 
varying CO emission profile and spectro-astrometric signature 
that we observe in HD~100546, future CO observations of this source 
would not be expected to show coherent velocity structure, e.g., 
such as that arising from orbital motion. 
If they arise from the effects of disk turbulence, further 
observations of variations in the CO emission from disks 
may help to provide evidence for, or constrain the nature of, 
disk turbulence. 

Another possibility is that the excess emission observed in
2006 and 2010 arises from substructure in the disk induced by the presence of an 
embedded companion.  Clarke \& Armitage (2003) have discussed the possibility that a gas giant planet forming in 
the disk around an FU Orionis object would result in 
non-axisymmetric circumstellar CO emission as material accreted onto the 
planet at a very high rate. Perhaps a similar result could arise from a forming gas giant planet in the transitional disk 
surrounding HD~100546. 

Substructure in the HD~100546 disk has been reported by Quanz et al. (2011) based on polarimetric differential 
imaging in $H$ and $Ks$ at 0$\arcsec$.1 resolution.  Their data, acquired in April 2006 (about 3 months after our 
2006 data were acquired) reveal a deficit (what they call a ``hole'') in the polarized flux and fractional polarization in 
both bands at $\sim$0$\degr$ east of north (see their figure 11), i.e., at the same position angle at which we posit an 
excess source of CO emission based on our 2006 data (Fig. 9).  Quanz et al. (2011) locate their ``hole'' at a radius 
of $\sim$27~AU from the star, further away than the inner radius of the disk.  While they note that it is difficult to 
determine whether their deficit is due to a lack of scatterers (dust grains) or a change in the scattering properties of 
grains at this position, they discuss how such a feature might be created as a shadow of a planet forming at smaller 
radii (Jang-Condell 2009), as one interpretation of their results.

There are a number of other lines of circumstantial evidence that suggest HD~100546 harbors a substellar 
companion. Firstly, transition disk SEDs such as the one associated with HD~100546 can be formed in response to 
dynamical sculpting by an embedded planet  (e.g. Skrutskie et al. 1990; Marsh \& Mahoney 1992; Bryden et al. 
1999; Calvet et al. 2002; Rice et al. 2003;  Quillen et al. 2004; D'Alessio et al. 2005; Calvet et al. 2005; Furlan et al. 
2006). Indeed Kraus \& Ireland (2012) have recently presented what may be the first images of a forming gas 
giant embedded in the transitional disk surrounding LkCa~15. Modeling of the SED led \citet{2003A&A...401..577B} 
to suggest that the transitional nature of HD~100546 may indicate the presence of a companion with a mass $
\gtrsim$5.6~M$\rm_{Jupiter}$. 

Secondly,  \citet{2005ApJ...620...470G} used long slit images acquired with the 
$Space\ Telescope\ Imaging\ Spectrograph$ on the $Hubble\ Space\ Telescope$ to map the inner edge of the disk. 
These authors found that the radius of the inner hole was 13$\pm$3~AU. They also noted that the star was offset 
from the center of the inner hole of the disk by $\sim$5~AU to the northeast. They concluded that such an offset was 
most readily explained by tidal interactions with an unseen massive companion. Similarly, in Paper I we measured 
rovibrational OH emission from the system and found that the observed asymmetry in the emission lines could be 
explained if the OH emission arose from a narrow annulus with $e\geq0.18$ at the inner edge of the disk. Such 
eccentricities are predicted to result from planet disk interactions (Papaloizou et al. 2001; Kley et al. 2006; Reg{\'a}ly et 
al. 2010). 

Thirdly,  evidence for a massive companion comes from the radial distribution of gas in the inner disk. The 
molecular gas is truncated at the same radius as the dust \citep{2009ApJ...702...85B, 2009A&A...500.1137V}. The 
suppression of the flow of material from the outer disk to the inner disk is expected if the clearing of the inner disk is 
due to an embedded companion (e.g. Lubow \& D'Angelo 2006). However, the companion is not massive enough 
to shut off accretion from the outer disk: the inner disk appears to be filled with \ion{O}{1} \citep{2006A&A...449..267A}
 and HD~100546 shows high velocity redshifted Balmer absorption lines indicative of ongoing stellar 
accretion \citep{2006A&A...457..581G}.  Finally, the generation of small dust grains in the innermost part of the disk 
may be indicative of dynamical stirring as may be expected with the presence of a massive companion \citep
{2010A&A...511A..75B}. 

How might an embedded companion give rise to the observed CO emission variability? One possibility is that the 
emission arises from a {\it circumplanetary disk}. 
Models of gas giant formation indicate that during the gas accretion 
phase, a circumplanetary disk roughly a third the size of the Hill sphere forms (Quillen \& Trilling 1998; Ayliffe \& 
Bate 2009ab; Martin \& Lubow 2011a). For a 5M$_{\rm Jupiter}$ planet forming 13~AU from a 2.4M$_\sun$ star, this 
corresponds to a disk radius of 0.4AU. CO emission from only a fraction of this area ($\sim 0.1\rm AU^2$) is needed to 
account for the flux excess we observe (sec. 3.3).  The narrow line profile of the excess CO line in 2006 (Figure 8, 
lower spectrum in red) is readily explained if the line 
originates in a relatively isothermal disk. In this case, the emission will be dominated by the outer disk which has an 
orbital velocity of $\sim$3~km~s$^{-1}$ at 0.3~AU.  The excess in the CO line in 2010 (Fig 8, lower spectrum in 
cyan) is different from that in 2006:  it is broader and double-peaked.  If these excesses arise in a disk, the 
difference in the profiles suggests that there was a significant change in the disk radial structure between these two 
epochs.

Circumplanetary disks are indeed expected to undergo outbursts (e.g. Martin \& Lubow 2011b; see also Zhu, 
Hartmann, \& Gammie 2009 and Armitage, Livio, \& Pringle 2001). Martin \& Lubow (2011b) describe how 
circumplanetary disks may undergo accretion in a non-steady state fashion through gravo-magneto instabilities. In 
other words, as the circumplanetary disk fills by accretion from the circumstellar disk, the surface density may rise 
sufficiently that the gravitational instability is triggered and the disk accretes vigorously onto the planet. Eventually 
the surface density drops sufficiently that a magneto-rotational instability may drive accretion. It is conceivable that 
the change in the line profile is a consequence of the changing emitting area as the disk fills and empties.
For example, an impulsive accretion event may heat the inner disk relative to the outer disk, producing 
a line profile that emphasizes smaller radii and larger rotational velocities. This would lead to a broader 
line profile and likely a larger line flux.

Variations in the accretion rate through the circumplanetary 
disk may also result from the eccentricity induced in the disk 
by a massive planet. 
In their study of the response of an accretion disk to a perturbing 
embedded planet, Kley \& Dirksen (2006) found that above a minimum 
planetary mass of q=0.003 times smaller than the stellar mass, the disk in the vicinity of 
the planet makes a transition from a nearly circular state to an 
eccentric state. The eccentricity is approximately stationary in 
the inertial frame.  In a companion paper (Liskowsky et al.\ 2012), 
we present evidence for such an eccentricity in the HD~100546 disk 
based on the line profile of the NIR OH emission. 

Kley \& Dirksen (2006) find that the eccentricity of the inner 
edge of the gap induced by the planet is such that when the planet is 
more massive than q=0.005, the disk is eccentric enough 
that the planet periodically passes through the gap edge.  
During these passages, the mass accretion rate onto the planet is 
strongly increased, allowing continued accretion onto the planet.
The accretion rate toward the planet may be more strongly pulsed if 
the orbit of the planet also becomes eccentric (D'Angelo, Lubow, \& Bate 2006). 

As discussed by Kley \& Dirksen (2006), 
the role of disk eccentricity in enhancing the planetary accretion 
rate may be important in allowing the growth of very high
mass planets (q$\sim 0.1$). 
At lower planetary masses where the disk is essentially circular, 
the width of the gap cleared by the planet grows as the planet grows in 
mass, eventually shutting off accretion onto the planet and star
(e.g., Bryden et al.\ 1999; Lubow et al.\ 1999; Lubow \& D'Angelo 2006),  
potentially making it difficult to form planets $\gtrsim 10 M_{\rm J}$.
Thus, it would be important to obtain evidence of eccentric disks 
with embedded planets and a varying circumplanetary accretion rate 
in order to confirm this picture.

Further observations of the HD~100546 system can test the hypothesis
of an orbiting companion. 
If the variations in the CO emission strength and spectroastrometric 
signal are due to the presence of an orbiting massive planet located 
just within the optically thick region of the disk (i.e., at $\sim 13$\,AU), 
we would expect to see the excess emission should shift blue-ward
over the next 5-7 years before disappearing behind the near side
of the disk for the next 15 years. The spectro-astrometric signal
on the blue side of profile would be reduced in spatial extent by
the presence of a compact line emission region with a similar
(blueshifted velocity) that is located close to the source of the
continuum emission. The asymmetric OH line profile reported in 
Liskowsky et al.\ (2012), if it arises from an eccentric inner disk 
rim induced by a massive orbiting companion, would remain 
approximately unchanged on this time scale since it  would be 
approximately stationary in the inertial frame and is expected to precess at a rate of 
only 10$\degr$ in 1000 orbits.

Reg{\'a}ly et al. (2010) considered the dynamical impact of a forming giant planet on the disk in which it is embedded 
and predicted the line profile variability that would be observed in the CO emission produced by the disk.  Their line 
profile predictions are less directly applicable to HD~100546, firstly because in their model the CO emits within, at, 
and beyond the orbit of the planet, whereas HD~100546 shows a clear inner hole to the CO emission at 13~AU.  
Whereas Reg{\'a}ly et al. (2010) did not consider the possibility of CO emission from the orbiting object itself, our 
results also indicate the possibility that significant CO emission may be produced by a compact orbiting source, 
possibly the circumplanetary disk of a forming planet.  Despite these differences, other aspects of the HD~100546 
system do resonate more with the picture described by Reg{\'a}ly et al. (2010).  For example, they confirmed the 
finding of Kley \& Dirksen~(2006) that an embedded planet induces a significant eccentricity in the disk.  As shown 
in Paper~I, we can reproduce the OH line profile from HD~100546 if it arises mainly from the inner wall of such an 
eccentric disk.

\section{Conclusions}%section 5

We have explored the use of multi-epoch spectroastrometry
and line profile asymmetries of CO fundamental emission
as probes of disk substructure in HD 100546.
For the higher vibrational CO lines controlled by UV fluorescence,
the spectro-astrometric signal at each of the four epochs is
consistent with emission from gas in an axisymmetric disk.
The equivalent width of the UV fluoresced lines vary, which we
argue is best explained by variations in the dust continuum emission.
In contrast to the UV fluoresced lines, the v=1-0 CO lines show
pronounced variations in the spectro-astrometric signal,
in addition to changes in flux and spectral shape.
The spectro-astrometric signal of the P26 lines has the signature
of axisymmetric disk emission in 2003.
However, in 2006 and in 2010,
the spectro-astrometric signal is asymmetric,
showing an offset only on the blueshifted side of the line.
There is also an increase in the flux of the v=1-0
CO emission in 2006 and 2010 relative to 2003.
The flux, central velocity, and velocity width of this
excess emission differs in 2006 and 2010.

We constructed a model with a compact source of emission near
the inner edge of the disk that can account for the variations
in the v=1-0 P26 line and its spectro-astrometric signal.
The emission strength, velocity width, and velocity offset of the compact source are
fixed to that of the excess P26 emission. If we assume that the velocity 
offset is due to a source in orbit around the star at the inner disk edge, 
the observed velocity offset implies a spatial location for the excess 
emission. By adding this prescription for the excess emission to the 
axisymmetric disk emission model, the changes in the spectro-astrometric 
signal can be reproduced. Thus the properties of the excess CO emission 
are consistent with an orbiting body whose CO emission strength and width vary in time.

We suggest that the compact CO emission could arise from
a circumplanetary disk orbiting a giant planet
near the inner disk edge,
with a variable accretion rate onto the circumplanetary disk.
Some other recent results suggestive of an
orbiting giant planet near the inner disk of HD100546
are relevant to this idea.
Quanz et al. (2011) find a "hole" near the inner disk edge
in polarimetric differential images of the HD 100546
taken within 3 months of our 2006 spectrum. This substructure
is at the same position angle as required by our model for
the excess CO emission at that epoch.  In Paper I,
we showed that the asymmetric line profile of the
OH rovibrational emission arising from the disk could
be reproduced if the inner edge of the disk has an eccentricity
of 0.18 as predicted by models of a disk interacting with a massive 
orbiting giant planet
(Papaloizou et al. 2001; Kley \& Dirksen 2006; Reg{\'a}ly et al. 2010). 

While our interpretation remains speculative, further observations can test this hypothesis.  If the excess v=1-0 
emission is due to a circumplanetary disk, then the excess emission should shift blue-ward over the next 5-7 years 
before disappearing behind the near side of the disk for the next 15 years. The spectro-astrometric signal on the 
blue side of the profile would be reduced in spatial extent by the presence of a compact line emission region with a 
similar (blueshifted velocity) that is located close to the source of the continuum emission.  

High signal to noise spectroastrometric observations are thus timely. Confirmation of a circumplanetary disk will 
open new opportunities to study the final stages of gas giant planet formation. 

If the variation we observed instead arises from random fluctuations in the disk, e.g., as a result of 
spatial inhomogeneities produced by the magnetorotational instability, we would not expect to observe systematic 
variations of this kind.  If the variations we observe arise from the effects of disk turbulence, further observations 
of CO emission from disks may help to provide evidence for, or constrain the nature of, disk turbulence.

\acknowledgments
Based on observations obtained at the Gemini Observatory, which is operated by the Association of Universities for 
Research in Astronomy, Inc., under a cooperative agreement with the NSF on behalf of the Gemini partnership: the 
National Science Foundation (United States), the Science and Technology Facilities Council (United Kingdom), the 
National Research Council (Canada), CONICYT (Chile), the Australian Research Council (Australia), MinistŽrio da 
Cincia e Tecnologia (Brazil) and SECYT (Argentina). The Phoenix infrared spectrograph was developed and is 
operated by the National Optical Astronomy Observatory. The Phoenix spectra were obtained as a part of programs 
GS-, GS-2005B-C-2, GS-2006A-C-17, GS-2008B-Q-22, and GS-2010B-C-3. S.D.B. acknowledges support for this 
work from the National Science Foundation under grant numbers AST-0708899 and AST-0954811and NASA 
Origins of Solar Systems under grant number NNX08AH90G. Basic research in infrared astronomy at the Naval 
Research Laboratory is supported by 6.1 base funding. M.R.T acknowledges support for this work was performed 
under contract with the Jet Propulsion Laboratory (JPL) funded by NASA through the Michelson Fellowship 
Program.  JPL is managed for NASA by the California Institute of Technology.

\appendix
\section{Analysis of potential artifacts}
One concern is the possibility that artifacts in spectroscopic data
may mimic or modify a spectro-astrometric signal
\citep[][]{2008LNP...742..123W}. 
To account for such artifacts, one can difference the measurement of 
parallel and anti-parallel slit positions. We checked for such artifacts in 2008 by collecting hot band data of 
particularly strong unblended lines near 2143~cm$^{-1}$ (Fig. 7). Plotting the parallel and antiparallel signals, it is
clear that the signals agree well though 
the combination of these measurements improves the sensitivity. The pixel-to-pixel variation
along the continuum has a standard deviation of 1.9~$mas$ for the individual spectra. The combined spectrum has a standard 
deviation along the continuum of 0.9~$mas$. 

Because we do not have parallel and anti-parallel observations for most of the epochs presented in this work, we have
carefully modeled several sources of artifacts: variable seeing, a distorted PSF, and the misalignment of the slit 
with the PSF. Here we briefly summarize the effect of each of these artifacts.  If the seeing is variable and the 
disk is resolved, the ratio of the emission line to the continuum will vary
as the portion of the disk excluded from the slit varies. This affects the strength of the spectro-astrometric signal, 
but it will not change the shape of that signal. The range of seeing conditions spanned by our observations 
($0\arcsec.6\pm0\arcsec.2$) results in a variation of the line to continuum ratio of $\pm$10\%. 

We also checked the effect of an elongated 
PSF to simulate the effect of poor focusing. We found that in principle this could create an 
artificial spectro-astrometric signal, however, it would effect all the lines the same way, whereas we observe
significant variation of the P26 line and no significant variation of the CO hot band lines.
We convolved the acquisition image with our model of the disk to check the effect 
of the actual focus on the resultant signal and found that it 
is not significant at our level  ($\lesssim 1mas$).  

Finally, we considered the effect of miscentering of 
the slit with the PSF of the source. This can lead to line asymmetries 
as one side of the disk is sampled more than the other. It also leads 
to an asymmetric spectro-astrometric signal.  However, miscentering 
will cause the brighter side to have a larger offset than the fainter side, which is 
the opposite of what is observed in our data (Fig. 6bc). Additionally, the PSF would need to be offset by
a full slit width to produce an observable effect. Miscentering the star on the slit by this amount is unlikely.

There is also some concern that telluric lines can introduce significant artifacts if the 
spectrum is not properly rectified. Inspection of the columns overlapping telluric lines (e.g. 2144.5 - 2145.5~cm$^{-1}$ in figure 7), the 
standard deviation is slightly higher (2.1~$mas$ and 2.2~$mas$ for the parallel and anti-parallel slit positions 
respectively). The standard deviation of the combined spectra is 1.1~$mas$, thus the improvement in
the sensitivity of our measurement of the centroid of the PSF is not necessarily more significant 
near the telluric features. Additionally, there are no apparent systematic artifacts that mimic the variation of the 
spectro-astrometric signal observed for the P26 line in 2003, 2006, and 2010 (Fig. 5). The spectro-astrometric 
signal of the hot band lines is consistent over all four epochs and the emission of the brightest 
emission lines can be seen to be spatially resolved (Fig. 4; see also Brittain et al. 2009). Thus we are confident that the 
spectro-astrometric signals we detect are not artifacts.

{\it Facilities:} \facility{Gemini:South (PHOENIX)}.

\clearpage
%%%%%%%%%%%%%%
% TABLE 1 - Observations %
%%%%%%%%%%%%%%
\begin{table}
\begin{center}
\caption[HD~100546: Log of Observations]{Log of Observations}\label{tab:observations}
\begin{tabular}{lccccccc}
	\hline
	Date  & Integration & Spectral Grasp & Slit PA & S/N & Seeing & Centroid rms & Airmass \\
	&  minutes & cm$^{-1}$ & deg E of N & & arcsec & milliarcseconds & sec(z) \\
	\hline \hline
	2003 January 7	 & 20 & 2029 - 2040 &  90 & 62  & 0.8 & 3.3 & 1.5 \\	
	2006 January 14	 & 12 & 2027 - 2038 &  90 & 43  & 0.6 & 4.3 & 1.4 \\
					 & 20 & 2141 - 2152 &  90 & 102 & 0.4 & 3.1& 1.3 \\
	2008 December 21   & 12 & 2136 - 2147 & 37 & 113 & 0.4 & 2.1 &  1.6 \\	
	2008 December 22   & 12 & 2136 - 2147 & 127 & 127 & 0.4 & 1.9 & 1.4 \\
					 & 12 & 2136 - 2147 & 307 & 122 & 0.4 & 1.9 & 1.6 \\ 		  
	2010 December 23 	 & 20 & 2027 - 2038 &  90& 79  & 0.8 & 4.5 & 1.4 \\
	\hline
\end{tabular}
\end{center}
\end{table}%

%%%%%%%%%%%%%%
% TABLE 2 - Clean Lines %
%%%%%%%%%%%%%%
\begin{table}
\begin{center}
\caption[HD100546: Unblended Lines]{Unblended UV Fluoresced Lines}\label{tab:cleanlines}
\begin{tabular}{lc}
	\hline
	Line & Wavenumber \\
	& cm$^{-1}$ \\
	\hline \hline
	v=3-2 P15 & 2030.16 \\
	v=6-5 R5 & 2032.83 \\
	v=4-3 P8 & 2033.14 \\
	$^{13}$CO v=1-0 P16 & 2033.42 \\
	v=3-2 P14	 & 2034.41 \\
	\hline
	v=2-1 R6 & 2142.47 \\
	v=3-2 R14 &  2142.72\\
	v=4-3 R23 &  2142.95\\
	$^{13}$CO v=1-0 R13 & 2144.01  \\
	\hline

\end{tabular}
\end{center}
\end{table}%

%%%%%%%%%%%%%%%%%%%%
% TABLE 3 - Equivalent Width of lines %
%%%%%%%%%%%%%%%%%%%%
\begin{table}
\begin{center}
\caption[HD100546: Equivalent Width of Average line profiles]{Equivalent Width of P26 and Averaged Hot Band Line Profiles}\label{tab:eqw}
\begin{tabular}{lll}
	\hline
	Date& Line & Equivalent Width \\
	& & 10$^{-2}$ cm$^{-1}$ \\
	\hline \hline
	2003 January 7	 	& P26            	& 4.50$\pm$0.14 \\	
					 	& HB (2032) 	& 1.63$\pm$0.12 \\
	2006 January 14	 	& P26            	& 6.46$\pm$0.15 \\	
					 	& HB (2032) 	& 1.85$\pm$0.13 \\
					 	& HB (2143) 	& 2.12$\pm$0.05\\
	2008 December 21-22 	& HB (2143) 	& 1.98$\pm$.010 \\				  
	2010 December 23	 	& P26 		& 9.80$\pm$0.13 \\	
					 	& HB (2032) 	& 2.50$\pm$0.12 \\
					
\hline

\end{tabular}
\end{center}
\end{table}%

%%%%%%%%%%%%%%%%%%%%
% TABLE 4 - Comparison of Excess     %
%%%%%%%%%%%%%%%%%%%%
\begin{table}
\begin{center}
\caption[HD100546: Comparison of excess Equivalent Width of P26 line ]{Properties of Excess CO v=1--0 Emission}\label{tab:excess_eqw}
\begin{tabular}{lllll}
	\hline
	Date& Scaled Equivalent Width$^1$ & Excess Equivalent Width & Doppler Shift of Excess & FWHM of Excess \\
	& 10$^{-2}$ cm$^{-1}$ & 10$^{-2}$ cm$^{-1}$ & km s$^{-1}$ & km s$^{-1}$   \\
	\hline \hline
	2003 January 7	 	& 4.50$\pm$0.14 & \nodata  & \nodata  & \nodata \\	
	2006 January 14	 	& 5.69$\pm$0.59 & 1.19$\pm$0.61 & +6$\pm$1 & 6 \\	
	2010 December 23	 	& 6.39$\pm$0.57 & 1.89$\pm$0.58 & -1$\pm$1 & 12 \\

\hline

\end{tabular}
\footnotetext{\footnotesize$^1$ The spectra were scaled such that the average hot band line profiles observed in 2006 and 2010 had the same equivalent widths as the average line profile observed in 2003.}
\end{center}
\end{table}%

%\clearpage

%%%%%%%%%%%
% FIG 1 - ALL Data %
%%%%%%%%%%%

\begin{figure}
    \begin{center}
	\includegraphics[scale=.65]{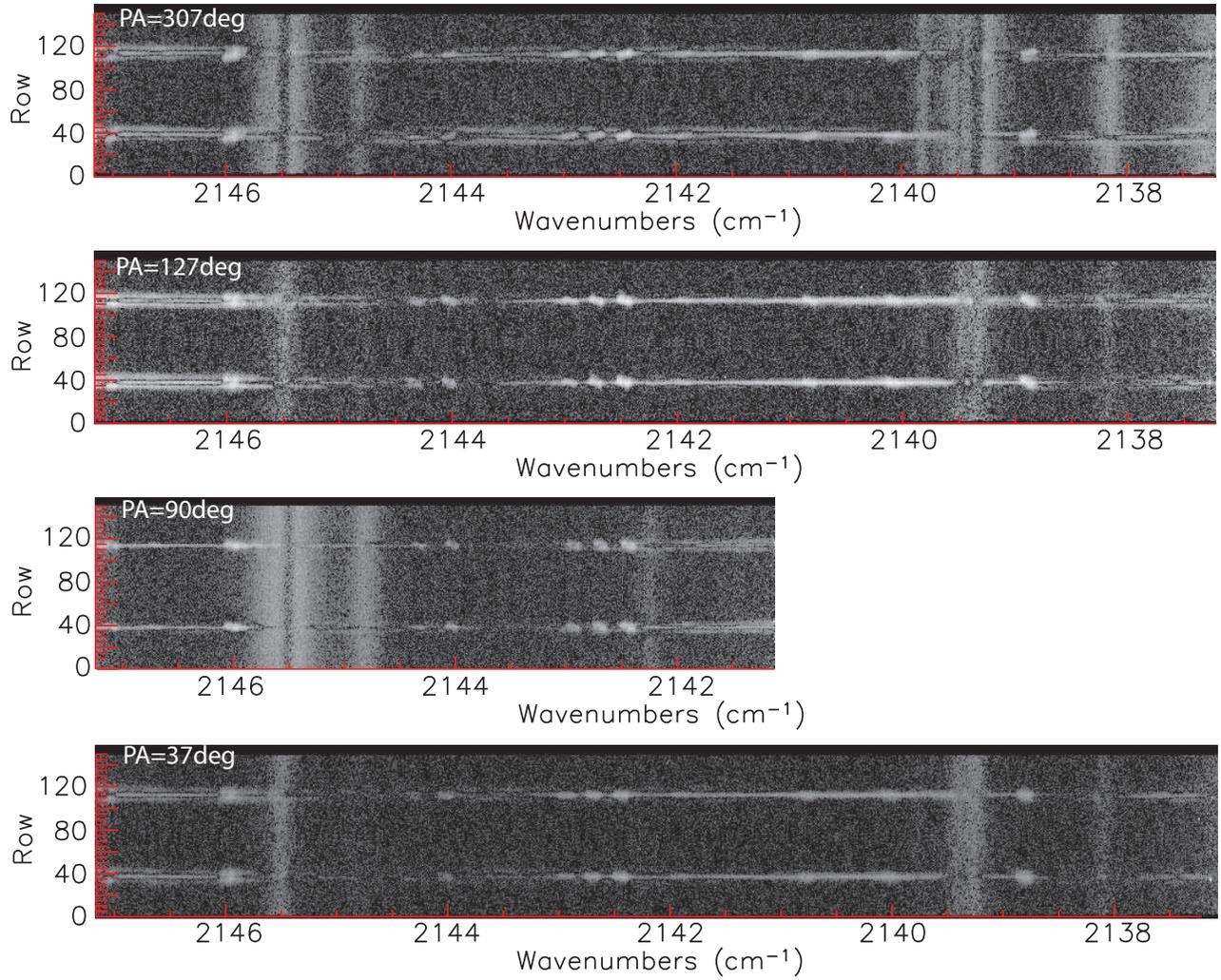}
	\caption[Spectroastrometry: 2008 Data]{Continuum subtracted images of the spectra centered near 
	2144 cm$^{-1}$. The average PSF was scaled using the telluric standard for each column. The 
	scaled PSF was then subtracted from the FITS image. These arrays were selected to show the effect 
	of rotating the slit on the spatial extent of the emission lines. Miscanceled sky lines appear as vertical 
	streaks in the data.}
	\label{fig:fig4}
   \end{center}
\end{figure}

\begin{figure}
    \begin{center}
	\includegraphics[scale=.6]{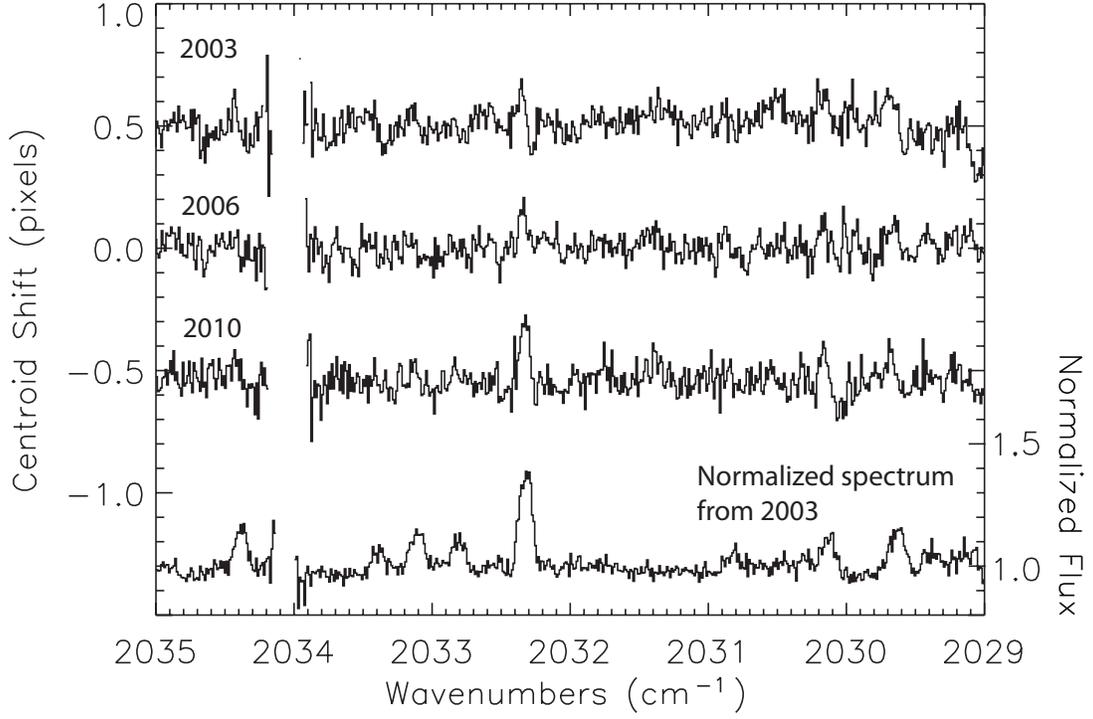}
	\caption[Spectroastrometry: 2008 Data]{Spectro-astrometric measurement of data centered near 
	2032 cm$^{-1}$. The centroid of each column is measured relative to the continuum and plotted in 
	pixels. The most prominent signal is from the $^{12}$CO v=1-0 P26 line. In 2003, it was symmetric as 
	is expected for gas arising from a Keplerian disk. In 2006 and 2010, the line was not symmetric as the 
	red side was no longer extended. Also plotted is the normalized spectrum acquired in 2003. The 
	strength of the spectro-astrometric signal scales with the line to continuum ratio of the feature which 
	varies.  The spectro-astrometric signal from the UV fluoresced lines is not seen individually for each 
	line because they are weaker than the v=1-0 P26 line. We measure this signal by averaging the 
	signals of the unblended lines (Fig. 6).   }
	\label{fig:fig5}
   \end{center}
\end{figure}

\begin{figure}
    \begin{center}
	\includegraphics[scale=.65]{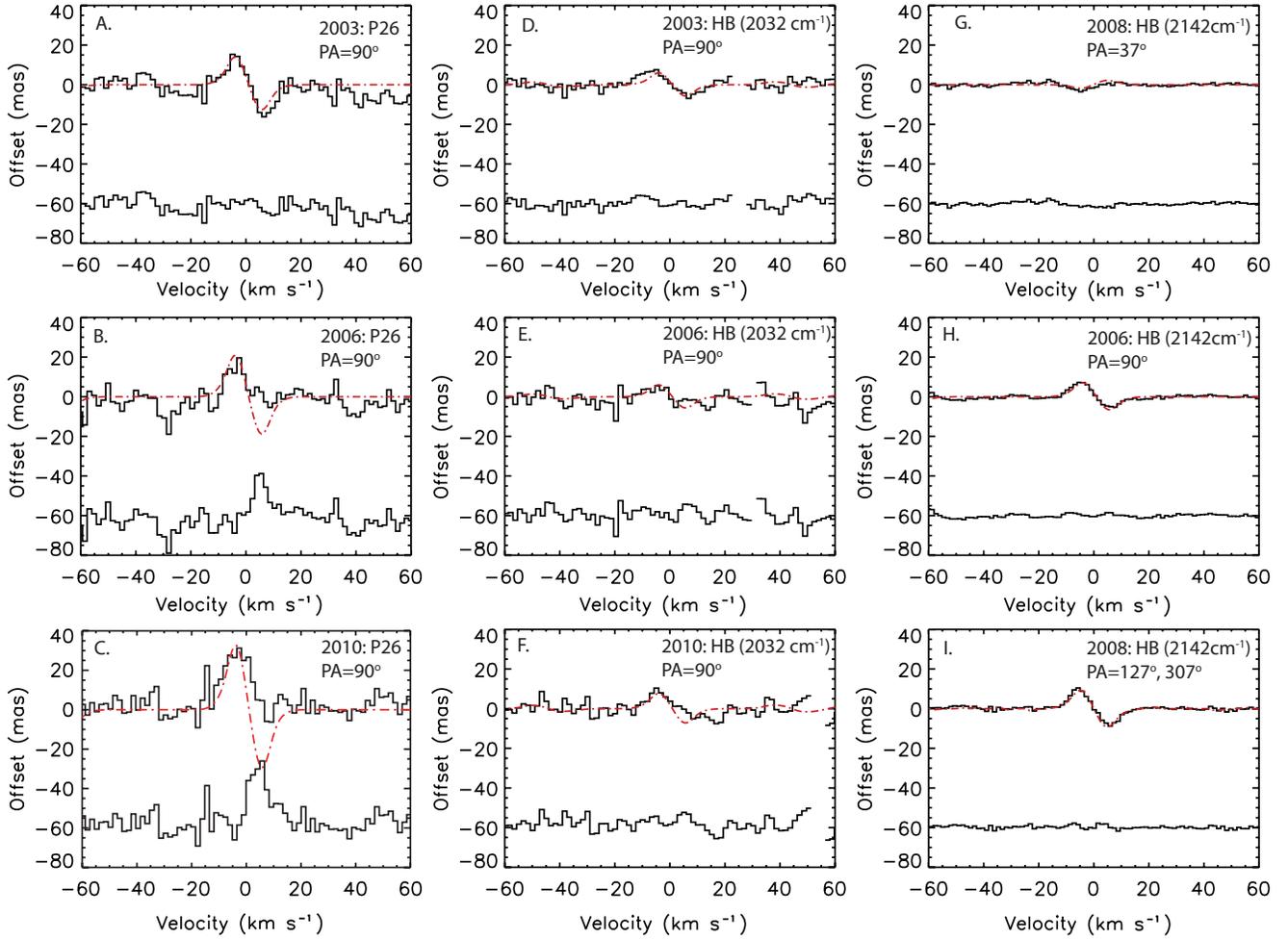}	
	\caption[Spectroastrometry: 2008 Data]{Spectroastrometric measurements of CO emission lines. The 
	shift in the center of light for each velocity channel of the spectrum is plotted (solid line above). For 
	each measurement, we plot the spectro-astrometric signal extracted from our fluorescence model 
	assuming the disk is axisymmetric (dot-dashed line). The difference between these is plotted below 
	(solid line). The extent of the shift depends on both the line to continuum contrast and spatial extent of 
	the emission. Because the hot band lines are significantly fainter than the v=1-0 $^{12}$CO lines, we 
	averaged the unblended hot band and $^{13}$CO lines to improve the S/N of the measurement 
	(panels d-i). The spectro-astrometric signal of the P26 line is also presented on each date (panels a-
	c). HD~100546 was also observed near 2142 cm$^{-1}$ in 2006 and 2008 (panels g-i). The hot band 
	lines observed near 2142 cm$^{-1}$ were observed with three position angles, 37$\degr$, 90$\degr
	$, and 127$\degr$. All other observations were acquired with the slit in its default position angle of 
	90$\degr$. The modeled spectroastrometric signal (dot-dashed line) of the emission lines is 
	calculated assuming the gas arises from a circular axisymmetric disk in Keplerian orbit. The hot band 
	lines are generally well fit (labeled HB). The P26 line is well fit in 2003 (panel a), but it is not fit in 
	2006 or 2010 (panels b and c). In the later epochs, the blue side of the emission line is spatially 
	extended, however, the red side is not. The center of light of the red side of the line is near the center 
	of light of the continuum point spread function.}
	\label{fig:fig7}
   \end{center}
\end{figure}

\begin{figure}
    \begin{center}
	\includegraphics[scale=.5]{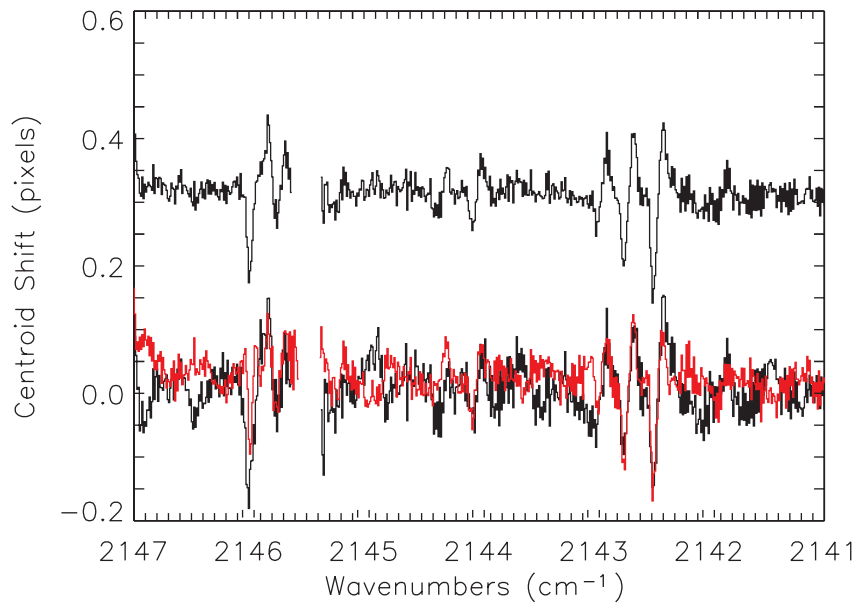}
	\caption[Spectroastrometry: comparison]{Spectroastrometric measurements of hot band CO 
	emission lines. The spectroastrometric measurement of the emission lines is measured with the slit 
	aligned nearly along the apparent semi-major axis (lower black line) and compared with an 
	observation taken in the anti-parallel position (red). The average of the two measurements is plotted 
	above. } 
	\label{fig:fig8}
   \end{center}
\end{figure}

\begin{figure}
    \begin{center}
	\includegraphics[scale=.5]{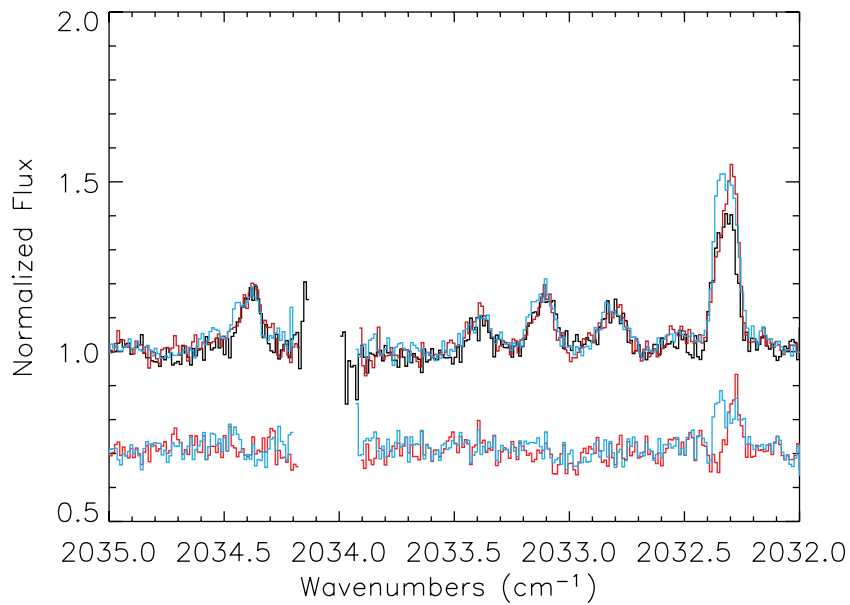}
	\caption[Spectroastrometry: 2008 Data]{Scaled and differenced spectra near 2033 cm$^{-1}$. The 
	spectra have been rescaled such that the equivalent width of the hot band lines is held constant over 
	all three epochs. The spectrum observed in 2003 (black) is subtracted from the spectra acquired in 
	2006 (red) and 2010 (cyan). The excess v=1-0 P26 emission is Doppler shifted 6$\pm$1 km s$^{-1}$ 
	in 2006 (red offset spectrum) while the v=1-0 P26 emission is shifted -1$\pm$1 km s$^{-1}$ in 2010 
	(cyan offset spectrum). In 2006 the emission is singly peaked and in 2010 the emission is double 
	peaked.}
	\label{fig:fig6}
   \end{center}
\end{figure}

\begin{figure}
    \begin{center}
	\includegraphics[scale=.5]{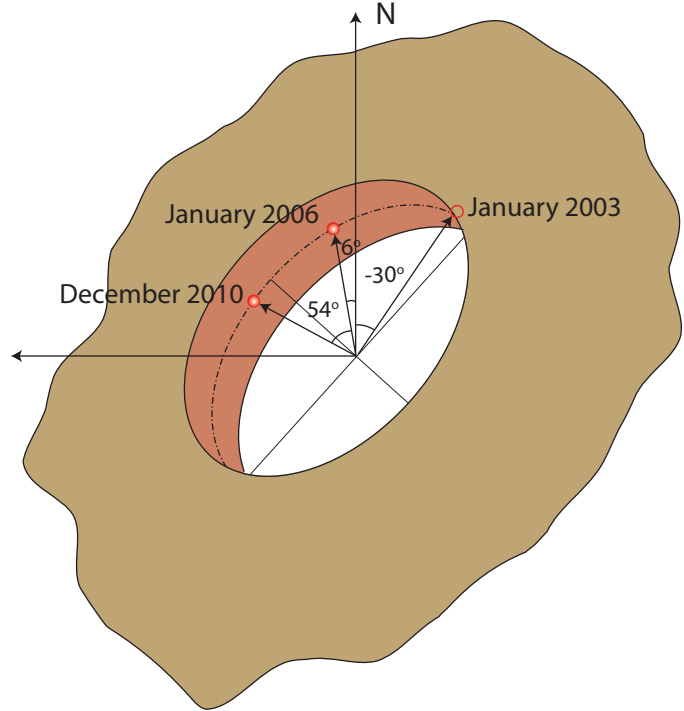}
	\caption[Schematic]{Schematic of geometry of observations. To reproduce the observed 
	spectroastrometric signal of the P26 line observed in 2006 and 2010, we posit an excess source of 
	emission in Keplerian orbit near the inner edge of the disk. The position angle of the emitting region 
	is $\sim$6$\degr$ E of N in 2006 and $\sim$54$\degr$ E of N in 2010. }
	\label{fig:fig9}
   \end{center}
\end{figure}

\begin{figure}
    \begin{center}
	\includegraphics[scale=.5]{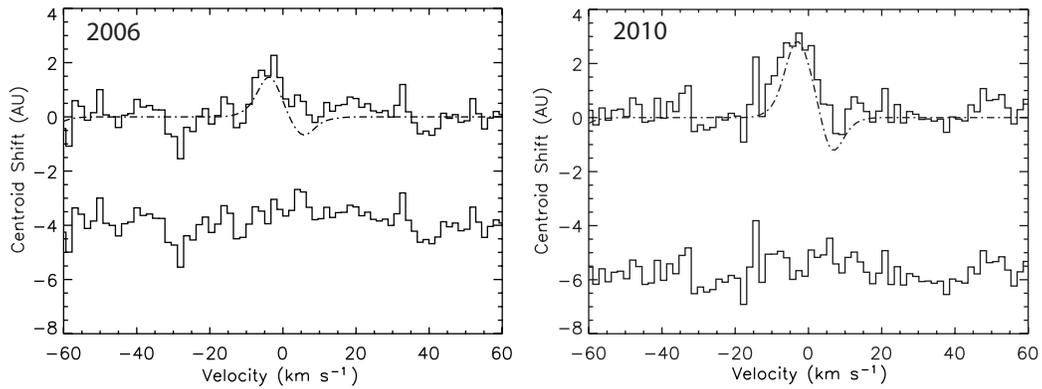}
	\caption[Spectroastrometry: comparison]{Spectroastrometric signal arising from the presence of a 
	non-axisymmetric source. By placing the excess emission at 6$\degr$ E of N and 54$\degr$ E of N 
	respectively, we can reproduce the spectroastrometric signal of our observations in 2006 and 2010 
	respectively.}
	\label{fig:fig10}
   \end{center}
\end{figure}

\end{document}